\documentclass[useAMS,usenatbib]{mn2e}

\usepackage{rotating}
\usepackage{multirow}

\usepackage{epsfig}
\usepackage{subfigure}

\usepackage{graphicx}
\DeclareGraphicsExtensions{.ps}

\makeatletter
\let\jnl@style=\rm
\def\ref@jnl#1{{\jnl@style#1}}

\def\aj{\ref@jnl{AJ}}                   
\def\araa{\ref@jnl{ARA\&A}}             
\def\apj{\ref@jnl{ApJ}}                 
\def\apjl{\ref@jnl{ApJ}}                
\def\apjs{\ref@jnl{ApJS}}               
\def\ao{\ref@jnl{Appl.~Opt.}}           
\def\apss{\ref@jnl{Ap\&SS}}             
\def\aap{\ref@jnl{A\&A}}                
\def\aapr{\ref@jnl{A\&A~Rev.}}          
\def\aaps{\ref@jnl{A\&AS}}              
\def\azh{\ref@jnl{AZh}}                 
\def\baas{\ref@jnl{BAAS}}               
\def\cjaa{\ref@jnl{ChJAA}}		
\def\jrasc{\ref@jnl{JRASC}}             
\def\memras{\ref@jnl{MmRAS}}            
\def\mnras{\ref@jnl{MNRAS}}             
\def\nar{\ref@jnl{NewAR}}               
\def\na{\ref@jnl{NewA}}                 
\def\pra{\ref@jnl{Phys.~Rev.~A}}        
\def\prb{\ref@jnl{Phys.~Rev.~B}}        
\def\prc{\ref@jnl{Phys.~Rev.~C}}        
\def\prd{\ref@jnl{Phys.~Rev.~D}}        
\def\pre{\ref@jnl{Phys.~Rev.~E}}        
\def\prl{\ref@jnl{Phys.~Rev.~Lett.}}    
\def\pasp{\ref@jnl{PASP}}               
\def\pasj{\ref@jnl{PASJ}}               
\def\qjras{\ref@jnl{QJRAS}}             
\def\skytel{\ref@jnl{S\&T}}             
\def\solphys{\ref@jnl{Sol.~Phys.}}      
\def\sovast{\ref@jnl{Soviet~Ast.}}      
\def\ssr{\ref@jnl{Space~Sci.~Rev.}}     
\def\zap{\ref@jnl{ZAp}}                 
\def\nat{\ref@jnl{Nature}}              
\def\iaucirc{\ref@jnl{IAU~Circ.}}       
\def\aplett{\ref@jnl{Astrophys.~Lett.}} 
\def\apspr{\ref@jnl{Astrophys.~Space~Phys.~Res.}}
\def\bain{\ref@jnl{Bull.~Astron.~Inst.~Netherlands}}
\def\fcp{\ref@jnl{Fund.~Cosmic~Phys.}}  
\def\gca{\ref@jnl{Geochim.~Cosmochim.~Acta}}   
\def\grl{\ref@jnl{Geophys.~Res.~Lett.}} 
\def\jcp{\ref@jnl{J.~Chem.~Phys.}}      
\def\jgr{\ref@jnl{J.~Geophys.~Res.}}    
\def\jqsrt{\ref@jnl{J.~Quant.~Spec.~Radiat.~Transf.}}
\def\memsai{\ref@jnl{Mem.~Soc.~Astron.~Italiana}}
\def\nphysa{\ref@jnl{Nucl.~Phys.~A}}   
\def\physrep{\ref@jnl{Phys.~Rep.}}   
\def\physscr{\ref@jnl{Phys.~Scr}}   
\def\planss{\ref@jnl{Planet.~Space~Sci.}}   
\def\procspie{\ref@jnl{Proc.~SPIE}}   

\begin{document}


\title{Sub-parsec radio cores in nearby Seyfert galaxies}
 

\author[Francesca Panessa and Marcello Giroletti]{Francesca Panessa$^{1}$\thanks{E-mail: francesca.panessa@iaps.inaf.it} and Marcello Giroletti$^{1}$\thanks{E-mail: giroletti@ira.inaf.it}\\
$^{1}$INAF - Istituto di Astrofisica e Planetologia Spaziali di Roma (IAPS), Via del Fosso del Cavaliere 100, 00133 Roma, Italy\\
$^{2}$INAF - Istituto di Radioastronomia, via Gobetti 101, 40129 Bologna, Italy}

\date{}

\maketitle
\begin{abstract}

We present a census of sub-pc scale properties of the VLBI cores in a complete sample of local Seyfert galaxies.
Seventeen out of 23 sources with a VLA detection are detected also with VLBI at 1.7 GHz and/or 5 GHz,
with an average monochromatic radio luminosity log [P$_\mathrm{5\, GHz}$/W Hz$^{-1}$] = 19.4.
Radio cores are of heterogeneous nature, the majority of them showing elongated structures or accompanied by extra components,
broad ranges of brightness temperatures (10$^{5}$-10$^{10}$ K) and spectral indices (from steep to highly inverted).
Interestingly, the detection rate (26\%) of water maser emission is considerably higher than that found in previous surveys ($\sim$ 10\%), suggesting that distance biases could significantly affect our knowledge of the actual occurrence of this phenomenon.
The VLBI observational properties of type 1 and type 2 nuclei are similar except for the T$_{B}$,
which is on average higher in type 1. These results suggest that both thermal and non-thermal emission
are common in low luminosity AGN, with a prevalence of free-free processes among type 2 cores, likely associated to molecular gas.
Though limited by the low number statistics, we find no significant correlation between the VLBI radio luminosity and the nuclear X-ray luminosity, 
the latter appears to be more connected to the tens of pc scales VLA radio emission, rather than to the sub-pc scales, particularly in the most X-ray luminous sources.
The X-ray radio-loudness parameter R$_{X}$ $\equiv$ L (6~cm)/L(2-10 keV), is on average very low ($\langle \log R_X \rangle$ = -4.8), with
comparatively higher R$_{X}$ found for sources with the largest black hole masses and the lowest Eddington ratios, although the radio power does not appear to depend on the accretion rate.

\end{abstract}


\section{Introduction}

Evidence of a correlation between radio and X-ray emission in X-ray binaries (XRBs, e.g., Gallo et al. 2003) and  Active Galactic Nuclei (AGN, e.g., Panessa et al. 2007, hereafter P07)
-- joined in the "fundamental plane" of black hole activity  (Merloni et al. 2003; Falcke et al. 2004) --
has suggested that the inflowing and outflowing components near the black hole are coupled, 
although the detailed physical/radiative mechanisms at work are matter of study.

Radio-quiet (RQ) AGN, like LINERs and Seyfert galaxies, emit radio waves at low flux levels (e.g., Ho \& Ulvestad 2001, Nagar et al. 2002).
However, their radio emission is weak and confined to arcsec scale, unlike in radio-loud (RL) AGNs, whose black holes are able to 
launch powerful relativistic jets up to kpc-Mpc scales. Yet, at the higher resolutions such as those mapped by the 
Very Long Baseline Interferometry (VLBI) images, significant compact radio emission is often found (e.g., Nagar et al. 2002, Anderson \& Ulvestad 2005, Wrobel \& Ho 2006). 
Even in the lowest luminosity objects, the radio emission is mostly unresolved at arcsec scale,
and often remains largely unresolved down to mas (Giroletti \& Panessa 2009, hereafter GP09). The origin of radio emission in RQ AGNs remains
still unclear and, when a stellar origin is excluded (Condon et al. 1991), 
it could be ascribed to a low-power jet (e.g., Miller, Rawlings \& Saunders 1993), to free-free emission from a molecular torus, from a disk wind
or the X-ray corona itself (Gallimore et al. 2004, hereafter G04). 

In this work, we present the first systematic study at VLBI spatial resolutions of a complete
sample of radio-quiet local Seyfert galaxies, with the purpose of characterizing statistically the physical
properties of the sub-pc cores and the incidence of associated jet/outflow structures.
In addition, we explore the relationship between the accretion related X-ray luminosity and the radio core luminosity, also
in dependence of their Eddington ratio and black hole mass.

In Section 1 we introduce the Seyfert galaxies sample and its Very Large Array (VLA) average properties.
The sub-parsec radio properties from VLBI observations are outlined in Section 2. In Section 3 we compare the radio luminosities versus the X-rays luminosities, black hole masses and Eddington ratios.
Finally, we discuss our results in section 4. Throughout this paper we assume a flat $\Lambda$CDM cosmology with ($\Omega_{\rm M}$,  $\Omega_{\rm\Lambda}) = (0.3$,0.7) and a Hubble constant of 70 km s$^{-1}$ Mpc$^{-1}$ (Bennett et al. 2003).

\section{The Sample}

The results here presented are based on a sample
of nearby Seyfert galaxies from the Palomar optical spectroscopic survey of nearby galaxies (Ho, Filippenko, \& Sargent 1995) complete to total B magnitude $B_T = 12.0$ mag.
In Cappi et al.\ 2006 (hereafter C06), the Palomar Seyfert sample has been
distance limited to $D \le 22$ Mpc and it consisted of 27 Seyfert
galaxies. Our final sample consists of 28 Seyfert galaxies (9 of type 1, 19 of type 2), including
the Seyfert 1.9 galaxy NGC\,3982, initially excluded by C06 because of the
lack of XMM-Newton data at that time, and it is distance limited to D$<$ 23 Mpc,
as NGC\,4639 has an updated measure of its distance to 22.9 Mpc.
No other sources in the Palomar survey have distances between 22 and 23 Mpc,
therefore the sample is still complete.
For a complete and detailed description of the sample here studied
we refer to C06 and references therein.

Most of the sources belonging to the sample are well studied local
Seyfert galaxies, observed with several radio  instruments at different resolutions.
In particular, observations of the nuclear region of all the sources in our sample were 
obtained by a systematic VLA radio survey of local nuclei conducted by Ho \& Ulvestad (2001). 
Data were taken at 6 cm and 20 cm with a $3\sigma$ threshold level of 0.12 mJy beam$^{-1}$
and an angular resolution of about 1$\arcsec$, which at the distances of our targets 
correspond to linear scales from 12.6 to 111 pc, for the nearest and farthest source respectively
(NGC\,4395 at D=2.6 Mpc and NGC\,4639 at D=22.9 Mpc). Eighteen out of 28 sources
have been detected at 20 cm (detection rate of $\sim$ 64\%) while only
5 sources were not detected at 6 cm ($\sim$ 82\% detection rate).
The typical VLA morphology shows a compact unresolved or slightly resolved core;
10/23 cores are accompanied by either diffuse emission or linear extended structures, 
five sources present a weak core defined as ambiguous; see 
Ho \& Ulvestad (2001) for a more detailed description of the cores morphology classification.
The extended structures reach up to 4.6 kpc scales. Nuclear spectral indices 
are equally distributed between steep and flat/inverted slopes, with thirteen sources showing $\alpha$ $<$ 0.5 (S $\propto$ $\nu^{-\alpha}$).

\section{The sub-parsec scale radio properties}

To have a census of the nuclear properties at higher spatial resolution,
we have collected VLBI data. Data for ten sources 
are already available in literature; typically these are bright sources 
with multiple observations available at different frequencies.
For the unobserved objects, which represent the majority of the sample,
we started an observational campaign with VLBI. In two recent papers (GP09, Bontempi et al. 2012, hereafter B12) we present deep quasi-simultaneous  dual
frequency (5 and 1.7 GHz) European VLBI Network (EVN) observations of fourteen objects.
Five sources (NGC\,3227, NGC\,3982, NGC\,4051, NGC\,4138 and NGC\,5033)
were detected at both frequencies, three more sources at only one frequency
(NGC\,4388 and NGC\,4501 at 1.7 GHz and NGC\,4477 at 5 GHz). Six 
sources were not detected at either frequencies at a level of $\sim$ 0.1 mJy. 

In Table~\ref{vlbi} we report the galaxy properties of the whole sample, 
together with data from VLBI and VLA measurements. The 2-10 keV X-ray luminosities here reported are corrected both for Compton thin and Compton thick absorption
as discussed in Panessa et al. (2006). In Table~\ref{stats} we present a statistical summary of the VLBI radio properties, in which
we computed average values including also upper limits; indeed, these results stem from  deep observations and therefore even non detections contribute a non-trivial information to the statistics. In that sense, the average values for the radio luminosity reported in Col. 13 are also upper limits. The same applies for the X-ray luminosities in Col. 12, which however have upper limits only for 2/28 sources.

Considering the observed sample, VLBI components at 5 GHz are found 
in 13 out of 21 nuclei, yielding
to a detection rate of $\sim$ 62\%; the detection rate for the type 1 and type 2 
classes are $\sim$ 75\% and $\sim$ 54\% respectively.
Similarly, the fraction of detected components at 1.6 GHz is $\sim$ 57\%,
with $\sim$ 78\% of detection rate for type 1 and 42\% for type 2.
Overall, $\sim$ 74\% have been detected at one frequency  at least.
Finally, we remark that five objects of the complete
sample (NGC\,676, NGC\,1058, NGC\,2685, NGC\,3486 and NGC\,4725, all of type 2) are undetected with VLA. 
Conservatively this yields to a detection rate of $\sim$ 61\% for the complete sample, 
although we cannot exclude the possible detection of a radio core with deeper observations.

An inspection of VLBI images reveals that the majority of the detected nuclei
consist of a single slightly resolved component, occasionally (in seven sources, including 1.6 GHz data) 
showing an elongated structure; only in the case of NGC\,1068, the core is characterized by diffuse disk-like
emission. One or more additional components are found in six sources. In this cases, we  focus on the feature that is most likely to be associated to the immediate vicinity of the central BH. Such feature is generally identified on the basis of its compactness and spectral properties, as discussed in the dedicated works (see Table 1, Col. 14  for the reference).
The VLBI core sizes are of the order of a few mas, where 1 mas corresponds to $\sim$ 0.05 pc
at the reference distance of 10 Mpc. At the low flux density of this sample, it is generally
hard to constrain the actual size of the components
and we conservatively treat these values as upper limits.

It is nonetheless interesting to estimate the brightness temperature $T_B$
in the present sample. We take $T_B$ values from the literature when available,
otherwise we directly calculate them as in B12. In the light of the above
mentioned caveat about the size of the components, we conservatively
consider all the resulting $T_B$ as lower limits, although it is well
possible that in most cases the actual brightness temperature is near the
estimated one. We further note that even if the observations come from
various different references, they are all overall quite similar in terms
of angular resolution, having been obtained with the VLBA and the EVN,
which have similar maximum baseline length (about 9000 km). The observing
frequency is also the same, since for most sources we have both 1.6 and 5
GHz data. 
Values range from $\sim$ 10$^{5}$ to $\sim$ 10$^{10}$ K. 
Ten out of 17 have T$_{B}$ $>$ 10$^{7}$ K which is typical of non-thermal emission (however see Blundell \& Kuncic 2007), among them type 1 and 2 classes are equally represented. 
Thermal processes are consistent with the data only in the remaining 7 sources, which are mainly of Type 2 class (5/7).
Thirteen sources have been observed at both 1.6 and 5 GHz, for these sources
which do not constitute an unbiased subsample, we can compute the spectral index,
the resulting values are broadly distributed between steep (including sources non detected at 5 GHz) and very inverted slopes,
with a prevalence of flat/inverted slopes. On average, type 2 nuclei have more flat/inverted spectral indices (see column 10 of Table~\ref{stats}).
The two optical classes do not show significant differences in the radio luminosity and the X-ray radio loudness parameter (R$_{X}$ $\equiv$ L(6~cm)/L(2-10 keV), Terashima \& Wilson 2003), with average values of log [L$_{R}$/erg s$^{-1}$] =  36.1 $\pm$ 1.0  and log R$_{X}$ = -4.8 $\pm$ 1.1.

At 5 GHz, the VLA over VLBI flux density ratios among the detected sources are broadly distributed between $\sim$1 and $\sim$100 
with no significant difference between type 1 and type 2 sources. In general, the majority of the VLA nuclear flux density
is resolved in jet-like complex sub-structures at VLBI resolution, as, e.g., in the Compton thick NGC\, 1068  (G04)
and in the type 1 NGC\, 3227 (B12). On the other hand, NGC\, 3031 (type 1.5, Marti-Vidal et al. 2011) and NGC\, 4565 (type 1.9, Falcke et al. 2000)
have a VLBI flux density larger than the VLA, likely due to intrinsic variability.

Of the 28 Seyfert galaxies in the sample, 27 have been observed at 22 GHz in search for H$_{2}$O maser emission (Zhang et al. 2012).
The typical water maser detection rate is of $\sim$7\% in AGN surveys (e.g. Braatz 
et al. 1997), with type 2 Seyfert galaxies having detection rates of up to 
$\sim$ 20\%, while in pure Sy 1 galaxies the fraction of water maser detections is always below $\sim$ 1\% (e.g. Braatz 
et al. 2004). Our complete sample, less biased by distance effects though sampling much smaller volumes than previous samples (recessional velocities $<$ 1200 km/s), 
reveals that the fraction of detected water maser emission is indeed significantly higher. Seven out of 27 ($\sim$ 26\%) have detected water maser 
emission, with an unusually large fraction of $\sim$ 22\% of type 1 objects detected, including the NLSy1 NGC\,4051 (see Tarchi et al. (2011) for a comparable detection rate in NLSy1). 
Interestingly, NGC\,5194 is one of the few known jet masers (Hagiwara et al. 2001), however our observations reveal that the radio jet is almost entirely resolved out at VLBI scales
(B12).

\begin{table*}
\scriptsize
\begin{center}
\caption{\bf The complete distance-limited sample of Seyfert galaxies\label{vlbi}}
\begin{tabular}{lrlrrrrrrrrrccr}
\hline
\hline
\multicolumn{1}{c}{Galaxy} &
\multicolumn{1}{c}{D} &
\multicolumn{1}{c}{Seyfert} &
\multicolumn{1}{c}{L$_{5}$} &
\multicolumn{1}{c}{L$_{1.7}$} &
\multicolumn{1}{c}{$\alpha$} &
\multicolumn{1}{c}{T$_{B, min}$} &
\multicolumn{1}{c}{L$_{5}$} &
\multicolumn{1}{c}{L$_{1.4}$} &
\multicolumn{1}{c}{$\alpha$} &
\multicolumn{1}{c}{L$_{X}$} &
\multicolumn{1}{c}{L$_{H2O}$} &
\multicolumn{1}{c}{M$_{BH}$} &
\multicolumn{1}{c}{L$_{X}$/L$_{Edd}$} &
\multicolumn{1}{c}{Ref.}\\
\multicolumn{1}{c}{Name} &
\multicolumn{1}{c}{} &
\multicolumn{1}{c}{Type} &
\multicolumn{1}{c}{VLBI} &
\multicolumn{1}{c}{VLBI} &
\multicolumn{1}{c}{VLBI} &
\multicolumn{1}{c}{VLBI} &
\multicolumn{1}{c}{VLA} &
\multicolumn{1}{c}{VLA} &
\multicolumn{1}{c}{VLA} &
\multicolumn{1}{c}{} &
\multicolumn{1}{c}{L$_{\odot}$} &
\multicolumn{1}{c}{M$_{\odot}$} &
\multicolumn{1}{c}{} &
\multicolumn{1}{c}{VLBI} \\
\multicolumn{1}{c}{(1)} &
\multicolumn{1}{c}{(2)} &
\multicolumn{1}{c}{(3)} &
\multicolumn{1}{c}{(4)} &
\multicolumn{1}{c}{(5)} &
\multicolumn{1}{c}{(6)} &
\multicolumn{1}{c}{(7)} &
\multicolumn{1}{c}{(8)} &
\multicolumn{1}{c}{(9)} &
\multicolumn{1}{c}{(10)} &
\multicolumn{1}{c}{(11)} &
\multicolumn{1}{c}{(12)}  &
\multicolumn{1}{c}{(13)}  &
\multicolumn{1}{c}{14} &
\multicolumn{1}{c}{15} \\
\hline
\hline
NGC 676  &    19.5  &	 S2:    &$-$	     &$-$        &$-$        & $-$	      &$<$35.61    & $<$34.92   &    $-$  &   40.79     &	 $-$    &      $-$		  &		  $-$		& -	 \\	
NGC 1058 &     9.1  &	  S2    &$-$   	     &$-$        &$-$        & $-$   	  &$<$34.78    & $<$34.23   &    $-$  &$<$37.55     &	 $<$ -0.1   &   4.9 	  &		 -5.43		& -    \\     
NGC 1068 &    14.4  &	S1.9    &   37.03    &$<$35.42   &$<$-2.25   &    6.6	  &   38.81    & 	38.64   &   0.69  &   42.84     &     2.2  &    7.2 		  &  	 -2.46		& 1	 \\ 
NGC 2685 &    16.2  &	 S2/T2: &$-$   	     &$-$        &$-$	     & $-$		  &$<$35.31    & $<$34.79   &    $-$  &   39.94     &     $<$ 0.4  &   7.1 		  &  	 -5.31	& -	\\     
NGC 3031 &     3.5  &	S1.5    &   36.92    &   36.27   &   -0.31   &    10.4    &   36.79    & 	  36.15 &  -0.16  &   40.25     &     $<$ -1.7	   &   7.8 	  &  	 -5.64			& 2	\\  
NGC 3079 &    17.3  &	  S2    &   37.35    &$<$35.04   & $<$-3.68   &    8.2	  &   38.22    & 	  37.57 &  -0.18  &   42.62     &     2.7	   &   7.7 		  &  	 -3.13		& 3	\\  
NGC 3185 &    21.3  &	 S2:    &$<$34.84	 &$<$34.22   &$-$        & $-$		  &   35.71    & $<$35.09   &$<$-0.12 &   40.79     &     $<$ -0.2	&   6.1 	  &  	 -3.37		& 4	\\  
NGC 3227 &    20.6  &	S1.5    &   36.16    &   35.97   &    0.62   &    7.5	  &   37.71    & 	  37.62 &   0.82  &   41.74     &     $<$ -0.2	&   7.6 	  &  	 -3.95			& 4	\\  
NGC 3486 &     7.4  &	  S2    &$-$   	     &$-$        &$-$        & $-$		  &$<$34.60    & $<$34.05   &    $-$  &   38.86     &   $<$ -0.2  &   6.1 		  &  	 -5.38	& -	\\     
NGC 3941 &    12.2  &	 S2:    &$<$34.87    &$<$34.49   &$-$        & $-$		  &   35.31    & $<$34.55   &$<$-0.37 &   38.88     &     $<$ 0.6  &   8.2 		  &  	 -7.37	& 4	\\  
NGC 3982 &    20.5  &	S1.9    &   36.34    &   36.02   &    0.37   &    7.6	  &   36.66    & 	  36.31 &   0.39  &   41.18     &    $<$ -0.4	&   6.1 	  &  	 -3.01			& 4	\\  
NGC 4051 &	  17  	&	S1.2    &   35.51    &   35.37   &    0.71   &    5.3	  &   36.57    & 	  36.32 &   0.55  &   41.31     &     0.3  &   6.1 			  &  	 -2.90		& 5	\\     
NGC 4138 &    13.8  &	S1.9    &   35.90    &   35.56   &    0.32   &    7.8	  &   35.95    & 	  35.21 &  -0.32  &   41.29     &    $<$ -0.5  &   7.6 		  &  	 -4.56		& 4	\\  
NGC 4151 &    20.3  &	S1.5    &   36.97    &   36.35   &   -0.25   &    8.3	  &    38.3    & 	  38.08 &   0.60  &   42.47     &     0.8  &   7.2 			  &  	 -2.81		& 6	\\  
NGC 4258 &     7.2  &	S1.9    &  $-$       &   34.34   &$-$        &    6.3	  &   35.74    & 	  35.18 &  -0.01  &   40.86     &     1.9  &   7.6 			  &  	 -4.85		& 7	\\  
NGC 4388 &    16.7  &	S1.9    &$<$35.94    &   35.82   &$>$0.77   &    6.1	  &   36.96    & 	  36.78 &   0.69  &   41.72     &     1.1	&    6.8 		  &  	 -3.18		& 5	\\     
NGC 4395 &     2.6  &	  S1    &   $-$      &   33.95   &$-$        &    6.3	  &   34.43    & 	  34.24 &   0.66  &   39.81     &     $<$ -1.4	&   5.0 	  &  	 -3.33			& 8	\\  
NGC 4472 &    16.7  &	S2::    &   36.54    &$-$     	 &$-$        &    7.5	  &   37.51    & 	  37.18 &   0.42  &$<$39.32     &     $<$ 0.4  &    8.8 	  &  	 -7.58			& 9	\\  
NGC 4477 &    16.8  &	  S2    &   38.35    &$<$34.08   &$<$-1.6    &    6.5	  &   35.46    & $<$34.76   &$<$-0.27 &   39.65     &     $<$ -0.2	&   7.9 	  &  	 -6.37	& 4	\\  
NGC 4501 &    16.8  &	S1.9    &$<$35.72    &   35.57   &$>$0.70   &    6.7	  &   36.25    & 	  35.94 &   0.44  &   39.59     &     $<$ -0.4	&    7.9 	  &  	 -6.41		& 5	\\     
NGC 4565 &     9.7  &	S1.9    &   36.23    &$-$	     &$-$        &    7.4	  &   36.17    & 	  35.51 &  -0.19  &   39.43     &     $<$ -0.7	&    7.7 	  &  	 -6.37			& 10	\\  
NGC 4579 &    16.8  &	  S1    &   37.53    &   36.95   &   -0.16   &    8.2	  &   37.81    & 	  36.95 &  -0.56  &   41.03     &     $<$ -0.4	&   7.8 	  &  	 -4.85			& 10,11\\  
NGC 4639 &    22.9  &	S1.0    &$<$35.17	 &$<$34.45   &$-$        & $-$   	  &   35.84    & $<$35.06   &$<$-0.41 &   40.22     &     $<$ 0.4  &   6.9 		  &  	 -4.73& 4	 \\  
NGC 4698 &    16.8  &	  S2    &$<$35.18    &$<$35.08   &$-$        & $-$   	  &   35.64    & $<$34.79   &$<$-0.54 &   39.16     &     $<$ 0.4 &    7.3 		  &  	 -6.24&  4  \\	   
NGC 4725 &    13.2  &	 S2:    &$-$   	     &$-$	     &$-$        & $-$   	  &$<$35.25    & $<$34.54   &    $-$  &   38.89     &     $<$ -0.5	&   7.5 	  &  	 -6.70		&  -  \\	      
NGC 5033 &    18.7  &	S1.5    &   36.18    &   35.64   &   -0.10   &    7.1	  &   36.80    & 	  36.53 &   0.51  &   41.08     &     $<$ -0.1  &      7.3	  &  	 -4.32			&   5 \\   
NGC 5194 &     8.4  &	  S2    &$<$34.80    &$<$33.98   &$-$        & $-$		  &   35.61    & 	  35.39 &   0.60  &   40.91     &     -0.2	&   7.0 		  &  	 -4.14	& 4 \\	  
NGC 5273 &    16.5  &	S1.5    &$<$35.34    &$<$34.65   &$-$        & $-$   	  &   36.22    & 	  35.93 &   0.49  &   41.36     &    $<$ 0.4  &   6.5 		  &  	 -3.25	&  5  \\	     
\hline													   
\hline
\end{tabular}
\\
Col. (1): galaxy name. Col. (2-3): distances in Mpc and optical classification as in Panessa et al. (2006),
Col. (4-5-6-7): log VLBI 5 GHz and 1.7 GHz luminosities in erg s$^{-1}$ (errors in the are conservatively estimated to
be 5-10\%, see VLBI references),  VLBI spectral index and log brightness temperature in K (T$_{B}$ are all lower limits); Col. (8-9-10):
log VLA 5 GHz and 1.4 GHz peak luminosities  in erg s$^{-1}$ (errors in the are conservatively estimated to
be $\sim$ 5\%, see Ho \& Ulvestad (2001), VLA spectral index as in Ho \& Ulvestad (2001); Col. (11): log 2-10 keV X-ray luminosity corrected for absorption (in erg s$^{-1}$) as in Panessa et al. (2006); Col. (12): log isotropic water maser luminosity in L$_{\odot}$ as in Zhang et al. (2012); Col: (13): log black hole masses in M$_{\odot}$ 
(see Panessa et al. 2006 for a discussion on uncertainties); Col. (14): Eddington ratio L$_{X}$/L$_{Edd}$; Col. (15): VLBI references: 1. G04; 2. Marti-Vidal et al. (2011); 3. Middelberg et al. (2007); 4. B12; 5. GP09; 6. Ulvestad et al. (2005); 7. Cecil et al. (2000); 8. Wrobel \& Ho (2006); 
9. Nagar et al. (2002); 10. Falcke et al. (2000); 11. Krips et al. (2007).
\end{center}
\end{table*}

\section{The nuclear X-ray versus radio comparison}

The hard 2-10 keV emission in Seyfert galaxies arises within the innermost region around the black hole
(i.e., the disk-corona system), while the diffuse extended components emit mostly in the soft band below 2 keV. In analogy with XRBs, the presence 
of a correlation between 2-10 keV and radio luminosity suggests an underlying disk-jet coupling mechanism (P07).

\begin{figure*}
\begin{center}
\parbox{16cm}{
\includegraphics[width=0.4\textwidth,height=0.3\textheight]{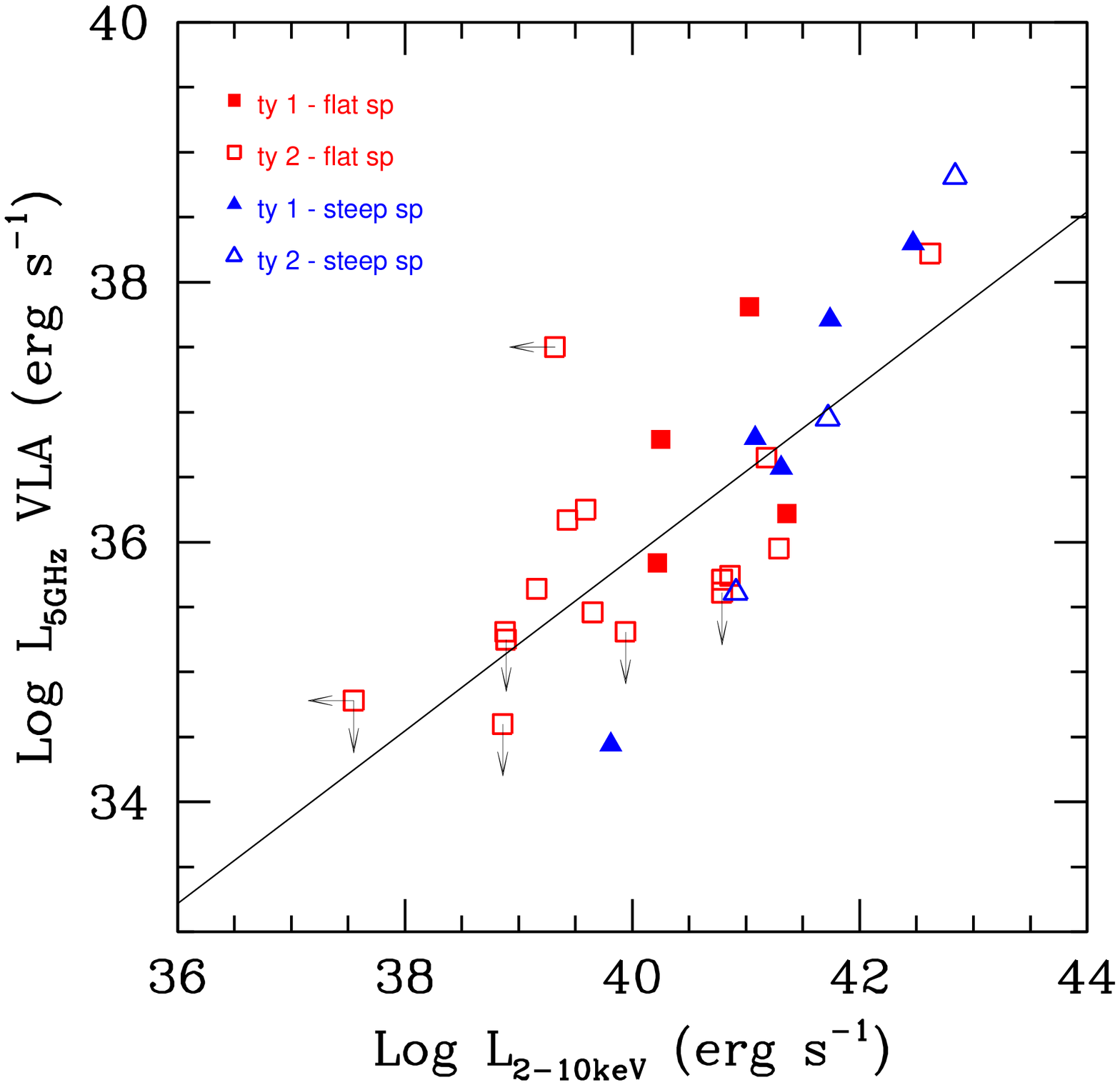}
\includegraphics[width=0.4\textwidth,height=0.3\textheight]{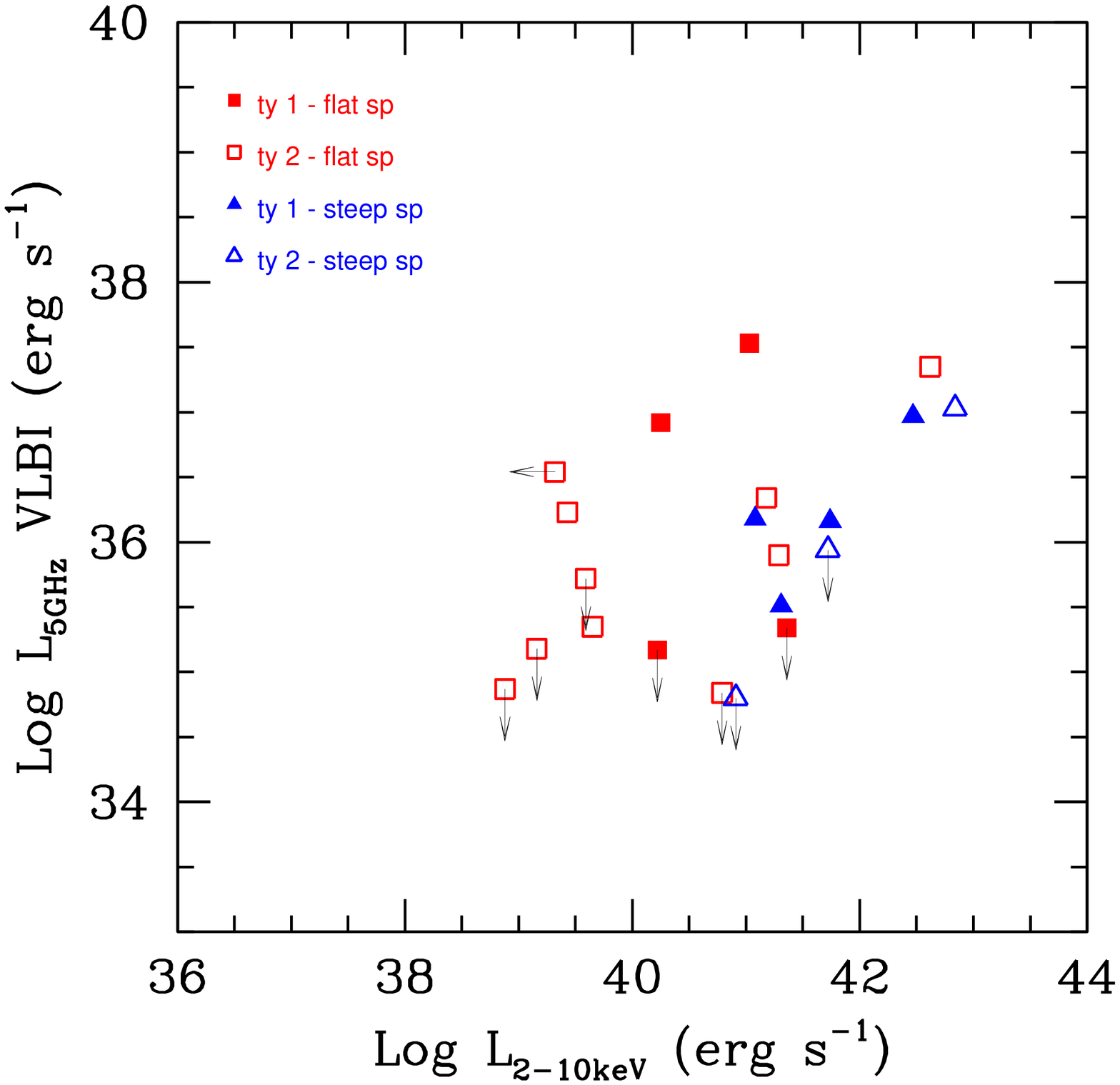}}
\caption{Radio luminosity at 5 GHz versus 2-10 keV unabsorbed luminosity. Left panel: VLA data.
Right panel: VLBI data. Type 1 and type 2 Seyferts are indicated with filled and empty symbols (see Panessa et al. 2006, for classification details), respectively; blue squares show flat spectrum radio sources, red triangles show steep spectrum ones, where spectral slopes are from VLA measurements.}
\label{vlbi_x}
\end{center}
\end{figure*}

To statistically characterize the relation between the radio versus X-ray data in our sample
we have applied the statistical survival analysis which takes into account censored data (ASURV package, Isobe, Feigelson \& Nelson 1986, Schmitt et al.  1985).
As dual censored points are present in our dataset, we have used the generalized Kendalls $\tau$ test for the correlation significance and the Schmitt's binned linear regression to calculate the slope of 
the correlation. In Figure~\ref{vlbi_x} we report the correlation for the distance limited sample, using
5 GHz luminosities measured with VLA (left panel) plotted versus the 2-10 keV unabsorbed luminosity.
A correlation between the VLA and X-ray luminosities is present. The Kendalls $\tau$ correlation probability P for accepting the null 
hypothesis that there is no correlation is 2.6$\times 10^{-5}$ ($z-value$ = 4.050, corresponding to a $\sim$ 4 $\sigma$ significance). 
Notwithstanding the fact that all sources are very nearby, we have also applied our tests to the flux-flux data, 
to rule out the 'distance-stretching' effect suffered by the luminosity-luminosity correlations,
and the correlation is still equally significant, with a probability of 3.7$\times 10^{-5}$  ($z-value$ =  3.964, $\sim$ 4 $\sigma$). 

The Schmitt's regression line slope is log [L$_{5 GHz}$/erg s$^{-1}$] $=$ 0.67 $\pm$ 0.01 log [L$_{2-10 keV}$/erg s$^{-1}$]. 
In both type 1 and type 2 AGN, the tens of pc scales radio emission seems to correlate with the nuclear activity, with a slope consistent with $\sim$ 0.7, as observed in low state XRBs (Gallo et al. 2003). On the other hand, though limited by the poor statistics, the 5 GHz VLBI mas scales radio core luminosities do not seem to correlate
with the X-ray emission, with a Kendall's $\tau$ probability of 0.05,  $z-value$ = 1.952, $<$ 1 $\sigma$ (Figure~\ref{vlbi_x}, right panel). 
This suggests that the X-ray emission correlates more significantly with the tens of parsec scale radio emission rather than with the nuclear sub-pc scales,
in agreement with recent findings in a sample of high energy selected AGN (Burlon et al. 2013, MNRAS accepted).
It is known that the non simultaneity of the radio and X-ray observations in AGN correlations introduces a significant source of scatter.
Indeed,  the brightest sources in the sample show strong X-ray variability (see C06), while the radio flux densities of most LLAGN show moderate
variability (Mundell et a. 2009), except for a few examples such as NGC\,3031. We expect the VLA and VLBI correlations to be affected in a similar way by flux variability.

\begin{figure*}
\begin{center}
\parbox{16cm}{
\includegraphics[width=0.4\textwidth,height=0.3\textheight]{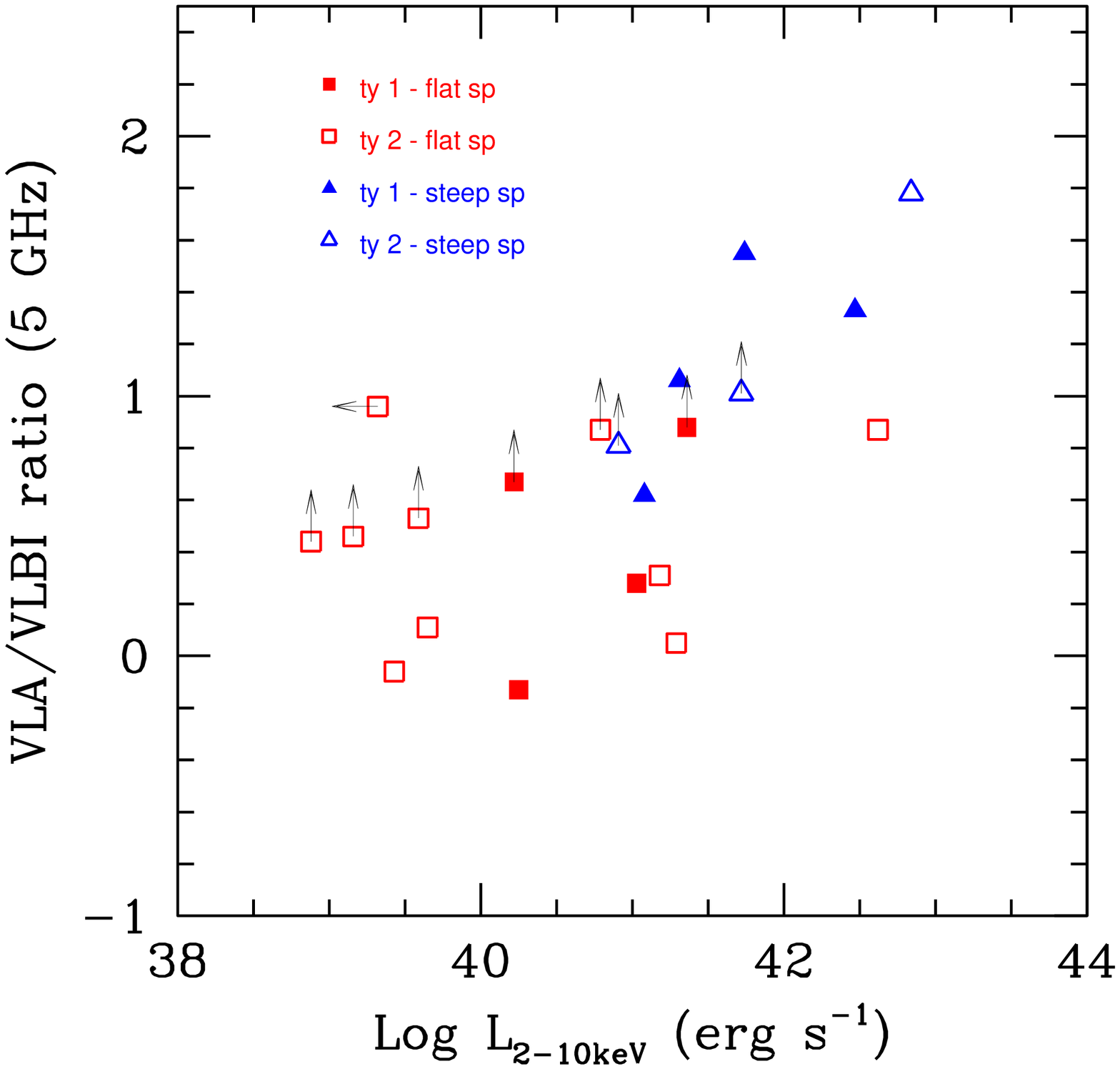}
\includegraphics[width=0.4\textwidth,height=0.3\textheight]{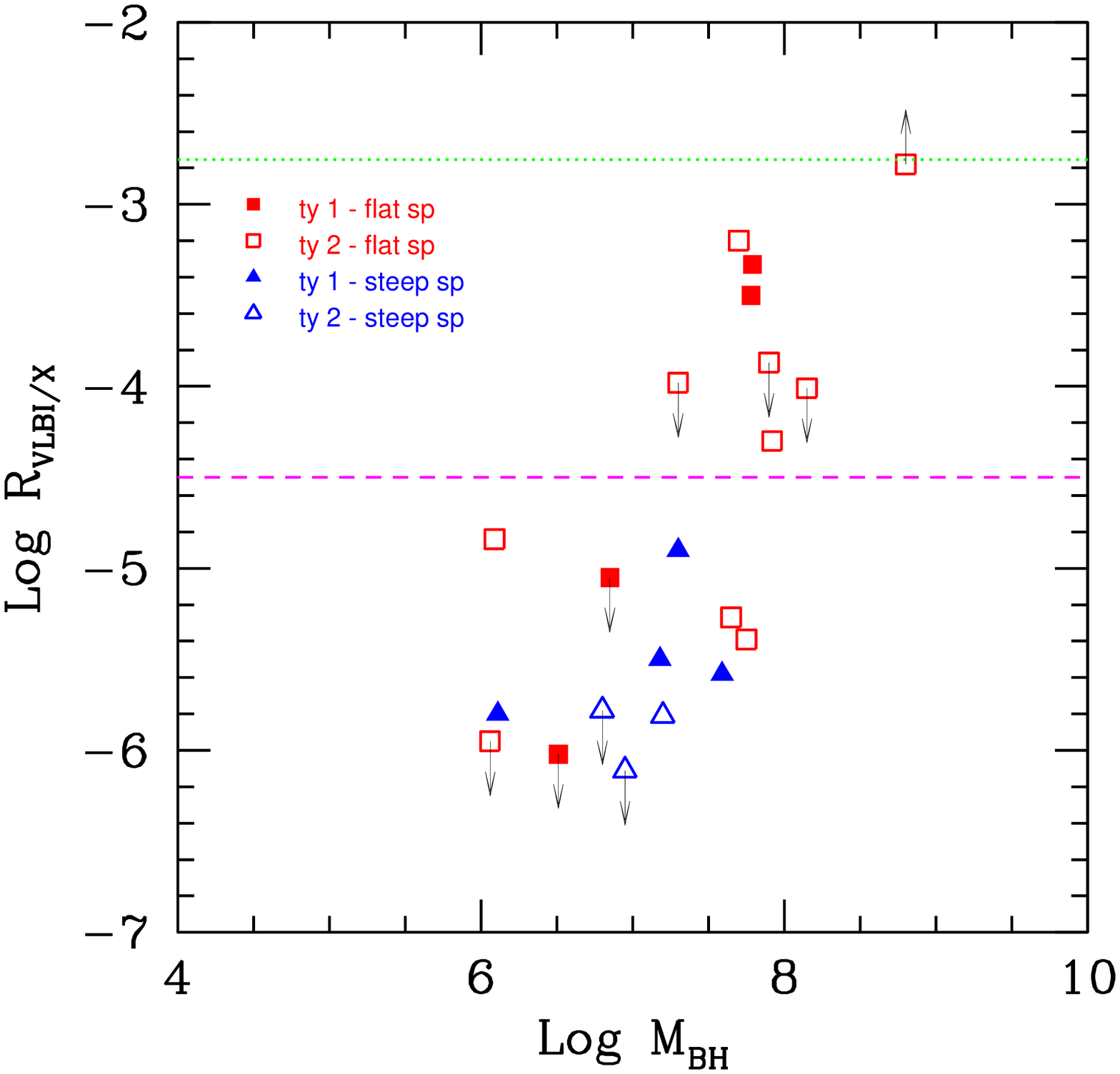}}
\caption{Left panel: VLA versus VLBI flux ratio versus 2-10 keV unabsorbed luminosity. Right panel: R$_{X}$ vs black hole mass.  
The magenta dashed line represent the R$_{X}$ $=$ -4.5
division between radio-loud and radio-quiet AGN (Terashima \& Wilson 2003), the green dotted line R$_{X}$ $=$ -2.755 has been derived in Panessa et al. (2006) for LLAGN. Symbols are as in Fig.\ 1.
\label{ra_x}}
\end{center}
\end{figure*}

Interestingly, at low X-ray luminosity,  most of the VLA flux is also measured with the VLBI.
While the more the source is X-ray bright, the more it has radio extended emission resolved at VLBI resolutions, as clear from Figure~\ref{ra_x}
(left panel) and, this effect is independent from the optical type (i.e., nuclear orientation).
Part of the diffuse/extended components emission could also contribute the hard X-ray luminosity (see 
e.g., NGC\,1068, NGC\,3079 and NGC\,4388,  heavily obscured X-ray nuclei embedded in extended soft diffuse emission).
On the other hand, unobscured type 1 AGN like NGC\,3227 and NGC\,4151 (X-ray point-like nuclei), placed in the top-right part of the right panel in Figure~\ref{ra_x}, 
suggest that the diffuse radio emission does not necessarily correlate with diffuse X-ray emission.
Finally, among the detected sources, clearly VLA steep sources tend to show larger VLA/VLBI ratios, while in flat cores
most of the VLA density flux is retrieved in the VLBI cores.

\begin{figure*}
\begin{center}
\parbox{16cm}{
\includegraphics[width=0.4\textwidth,height=0.3\textheight]{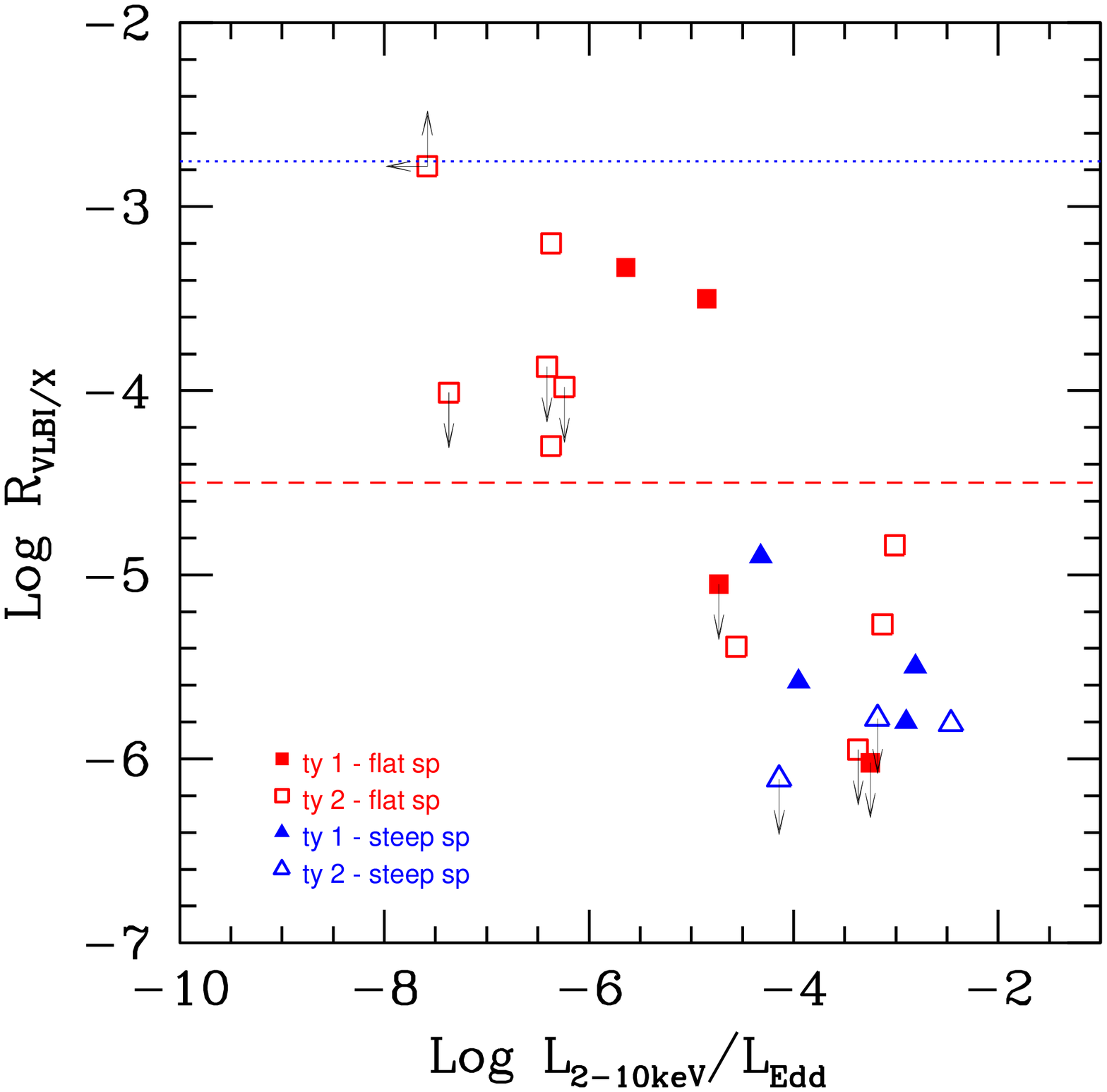}
\includegraphics[width=0.4\textwidth,height=0.3\textheight]{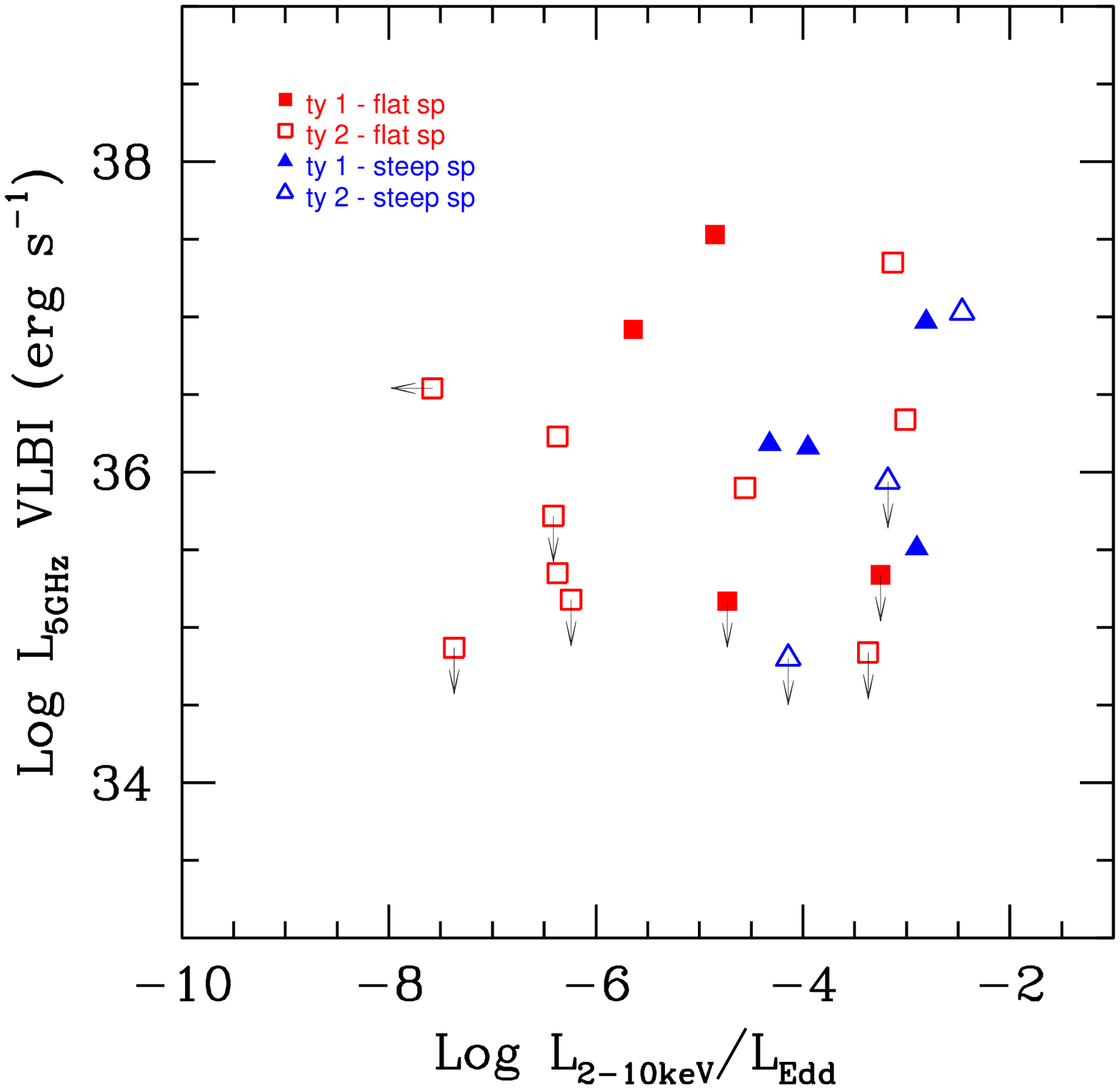}}
\caption{Left panel:radio loudness
parameter R$_{X}$ versus the Eddington ratio L$_{2-10 keV}$/L$_{Edd}$. Right panel: VLBI luminosity versus Eddington ratio. Symbols as in Fig.\ 1.}
\label{fp}
\end{center}
\end{figure*}

Though it is now established that the radio loudness parameter does not describe two necessarily different AGN radio populations (La Franca, Melini \& Fiore 2010),
the radio-loud vs radio-quiet terminology has still to be used as an indicator of the radio over the X-ray emission dominance.
The VLBI cores in this sample are, on average, extremely radio-quiet; five source 
are radio-loud (NGC\,3031, NGC\,4472, NGC\,4477, NGC\,4565 and NGC\,4579), according to the Terashima \& Wilson (2003)
limit (R$_{X}$ $=$ -4.5). If, instead, the P07 limit at R$_{X}$ $\sim$ -2.8 is applied, than all the nuclei in our sample are radio-quiet.
Indeed, even when compared to the least radio powerful VLBI cores of radio galaxies (Liuzzo et al. 2009), the Seyfert cores are still 2-3 orders of magnitude more radio-quiet. 
The only elliptical galaxy in our sample, NGC\,4472, is indeed the most radio-loud source.
In the right panel of Figure~\ref{fp} we show that, overall, all the radio-loud objects have: i) log [M$_{BH}$/M$_{\odot}$] $\geq$ 10$^{7}$, in agreement with the well known association
of RL objects with large M$_{BH}$; ii) the lowest Eddington ratios (calculated as the ratio between the unabsorbed 2-10 keV and the Eddington luminosities, the latter derived using M$_{BH}$ as in Table 1), below L$_{2-10 keV}$/L$_{Edd}$ $\sim$ 10$^{-4}$ (Sikora et al. 2007). 
However, the radio power does not seem to significantly correlates with Eddington ratio, 
as shown by the lack of any trend in the corresponding plot (shown in the right panel of Fig.~\ref{fp}).

\begin{table*}
\tiny
\begin{center}
\caption{\bf Summary of VLBI statistical results\label{stats}}
\begin{tabular}{lcccccccccccccc}
\hline
\hline
\multicolumn{1}{c}{Sample} &
\multicolumn{2}{c}{det.\ rate} &
\multicolumn{2}{c}{det.\ rate} &
\multicolumn{2}{c}{det.\ rate} &
\multicolumn{1}{c}{$\langle \log T_{B} \rangle$} &
\multicolumn{1}{c}{$T_{B}$} &
\multicolumn{1}{c}{$\langle \alpha \rangle$} &
\multicolumn{1}{c}{$\alpha$} &
\multicolumn{1}{c}{$\langle \log R_X \rangle$} &
\multicolumn{1}{c}{$\langle L_{5}^{\rm{VLBI}} \rangle$} &
\multicolumn{2}{c}{Maser} \\
&
\multicolumn{2}{c}{at either freq.} &
\multicolumn{2}{c}{1.7 GHz} &
\multicolumn{2}{c}{5 GHz} & 
\multicolumn{1}{c}{(K)} & 
\multicolumn{1}{c}{high:low} & & 
\multicolumn{1}{c}{flat:steep} & &
\multicolumn{1}{c}{(erg s$^{-1}$)} &
\multicolumn{2}{c}{det. rate} 
\\
\multicolumn{1}{c}{(1)} &
\multicolumn{1}{c}{(2)} &
\multicolumn{1}{c}{(3)} &
\multicolumn{1}{c}{(4)} &
\multicolumn{1}{c}{(5)} &
\multicolumn{1}{c}{(6)} &
\multicolumn{1}{c}{(7)} &
\multicolumn{1}{c}{(8)} &
\multicolumn{1}{c}{(9)} &
\multicolumn{1}{c}{(10)} &
\multicolumn{1}{c}{(11)} &
\multicolumn{1}{c}{(12)} &
\multicolumn{1}{c}{(13)} &
\multicolumn{1}{c}{(14)} \\
\hline
\hline
Full  & 17/23 & 74\% & 12/21 & 57\% & 13/21 & 62\% & $7.3\pm1.2$ & $10:7$ &$-0.2\pm1.4$ & $9:4$ & $-4.8\pm1.1$ & $36.1 \pm 1.0$ & 7/27 & 26\%\\
Type 1 & 7/9 & 78\% & 7/9 & 78\% & 6/8 & 75\% & $7.6\pm1.6$ & $5:2$ & $0.4\pm0.2$ & $4:2$ & $-5.0\pm1.0$ & $36.2 \pm 0.9$ & 2/9 & 22\%\\ 
Type 2 & 10/14 & 71\% & 5/12 & 42\% & 7/13 & 54\% & $7.1\pm0.7$ & $5:5$ & $-0.8\pm1.7$ & $5:2$ & $-4.7\pm1.1$ & $36.1 \pm 1.1$ & 5/18 & 28\%\\ 
\hline													   
\hline
\end{tabular}
\end{center}
\scriptsize
Threshold values for Cols.\ (9) and (11) are $\log [T_B/K]$ of 7.0 and $\alpha$ of 0.5.
\end{table*}

\section{Discussion and Conclusions}

The milliarcsecond radio cores of Seyfert galaxies in the local Universe display a variety of morphologies, power and emission properties.
The detection rate is likely around 60\% (up to $\sim$ 74\%), at 5 GHz. However,
we cannot exclude that the undetected nuclei could be weaker and detectable at higher sensitivities, nor can we exclude flux variability; 
as for the undetected type 1 nucleus NGC\, 5273, where VLA 8.4 GHz long term variability has been found (Mundell et al. 2009).
Such a detection rate for Seyfert galaxies is low when compared to the VLA and to the X-ray detection rates. Indeed, at mas scales, a detection rate of 100\% 
is reported for LLAGN only when flat spectrum radio cores are selected (Nagar et al. 2002, Falcke et al. 2000). 
However, the radio powers here sampled are very low and deeper observations are fundamental to establish the frequency of radio cores in weak AGN nuclei.

We find evidence or hints of the presence of a jet/outflow feature in the majority of the VLBI detected sources,
either extending up to kpc scales as in NGC\,1068 (G04), NGC\,3227 (B12), and
NGC\, 4258 (Cecil et al. 2000), or being confined to the inner pc/sub-pc regions 
(NGC\, 3079, NGC\,4151, NGC\,4395), possibly because of the jet interaction with the inter stellar medium (e.g., Ulvestad et al.  2005, Wrobel \& Ho 2006, Middelberg et al. 2007). 
In the case of NGC\,3031, a weak one-side jet is observed and
its extreme variability has been interpreted as caused by a precessing jet (Marti-Vidal et al. 2011). A double
($\sim$40 pc extended) lobe-like structure is observed in NGC\,4051, which resembles a mini radio galaxy (GP09).
In objects like NGC\,4395, the VLBI core has an associated elongated structure (0.3 pc) reminiscent of a possible outflow,
similarly as in NGC\,3227, NGC\,3982, NGC\,4138 and NGC\,4579 (B12, Falcke et al. 2000). 
Only in two sources the disk-jet orientation can be guessed assuming that the disk is traced by water maser emission:
the sub-pc jet in NGC\,4258 is orthogonal to the disk (Cecil et al. 2000), while in NGC\,1068
the core radio axis and the line traced by the water maser are slightly misaligned (G04).
Interestingly, the presence of a jet/outflow structure does not correlate with the core radio-loudness parameter,
as sources like NGC\,1068, NGC\,3079, NGC\,3227 and NGC\,4151 all have log R$_{X} <$ -5.5.

The origin of the physical/radiative mechanisms in radio cores and its coupling with X-ray emission has been the subject of a plethora of interpretations and models 
(e.g., Merloni et al. 2003, Falcke et al. 2004, Markoff et al. 2005, Ishibashi \& Courvoisier 2010).
High brightness temperature, flat spectrum compact radio cores 
are typically associated to non-thermal self-absorbed synchrotron (SSA) emission, as originating 
from the base of a jet; ten sources in our sample have high T$_{B}$,
eight of them have also an estimate of the spectral radio index: all but one (NGC\,3227)
have flat/inverted spectra supporting the SSA origin of their emission. 
Interestingly, type 1 cores show higher T$_{B}$ with respect to type 2 cores,
suggesting that the role of free-free processes associated to the presence of a dense molecular gas
is more prominent in type 2 Seyferts, in agreement with unified model predictions.
If the radio emission in low T$_{B}$ type 2 Seyferts
is associated to thermal free-free emission, the estimated electron density is of the order of 10$^{4-5}$ cm$^{-3}$ (B12, G04).
On the other hand, some low $T_{\rm{B}}$ cores, like NGC\,4051, can be of non thermal origin as suggested by the steep spectral index and the presence of aligned components
on parsec scales.

Laor \& Behar (2008) propose that both the radio and X-ray emission in radio-quiet QSO originates in coronal activity, with a predicted L$_{R}$/L$_{X}$ $\sim$ 10$^{-5}$
as observed in coronally active stars. Radio-loud cores exceed this ratio (as in Figure~\ref{ra_x}, right panel), requiring
an additional source of radio emission, likely associated to the base of a jet. Sources with L$_{R}$/L$_{X}$ $\sim$ 10$^{-5}$ are consistent with the idea of a hot corona origin 
of their radio emission. As an example, NGC\,1068 has a VLBI disk-like core associated with an 
X-ray hot corona or a wind arising from a molecular disk (G04).

The results presented in this work, though based on a well-defined sample, are limited by the heterogeneity of the VLBI observations
and by the non simultaneity of the radio versus X-ray observations.
Deep multi-frequency radio observations combined with simultaneous
X-ray data are needed (as in e.g. Bell et al. 2011, Jones et al. 2011), on a sizeable sample of AGN to investigate more in detail the physical processes at work.
In particular, the unprecedented sensitivity, spatial resolution and monitoring 
capabilities of the new generation Square Kilometer Array (SKA) will be extremely relevant in overcoming the present limitations.

\section*{Acknowledgments}

We acknowledge the anonymous referee for his/her suggestions. We thank Andrea Merloni, Andrea Tarchi, Loredana Bassani and Davide Burlon for discussions. The EVN is a joint facility of European, Chinese, South
African and other radio astronomy institutes funded by their national
research councils. This work has benefited from research funding from the EUFP6 RadioNet R113CT 2003 5058187 and the
FP under grant agreement No. 283393 (RadioNet3). FP acknowledges support from INTEGRAL ASI I/033/10/0, ASI/INAF I/009/10/0 and
the COST action "Black Holes in a Violent Universe".

\end{document}